# Redox-governed charge doping dictated by interfacial diffusion in two-dimensional materials

Kwanghee Park[1]†, Haneul Kang[1]†, Seonghyun Koo[1], DaeEung Lee[2] and Sunmin Ryu[1,3]*

[1] Department of Chemistry, Pohang University of Science and Technology (POSTECH), Pohang, Gyeongbuk 37673, Korea.

[2] Department of Applied Chemistry, Kyung Hee University, Yongin, Gyeonggi 17104, Korea.

[3] Division of Advanced Materials Science, Pohang University of Science and Technology (POSTECH), Pohang, Gyeongbuk 37673, Korea.

*Correspondence to: sunryu@postech.ac.kr

†They contributed equally.

**Abstract**

Controlling extra charge carriers is pivotal in manipulating electronic, optical, and magnetic properties of various two-dimensional (2D) materials. Nonetheless, the ubiquitous hole doping of 2D materials in the air and acids has been controversial in its mechanistic details. Here we show their common origin is an electrochemical reaction driven by redox couples of oxygen and water molecules. Using real-time photoluminescence imaging of $WS_2$ and Raman spectroscopy of graphene, we capture molecular diffusion through the 2D nanoscopic space between 2D materials and hydrophilic substrates, and show that the latter accommodate water molecules also serving as a hydrating solvent. We also demonstrate that HCl-induced doping is governed by dissolved $O_2$ and pH in accordance with the Nernst equation. The nanoscopic electrochemistry anatomized in this work sets an ambient limit to material properties, which is universal to not only 2D but also other forms of materials.

**Introduction**

Reduction in dimensions has seen many scientific discoveries in various metallic and semiconducting low-dimensional materials during the past four decades. Because of their high fraction of surface atoms, in particular, various material properties of two-dimensional (2D) materials are greatly affected by charge exchange with neighboring chemical entities or environments. When exposed to alkali metals [1] or halogens [2], for example, the Fermi level ($E_F$) of graphene swings by several tenths of 1 eV with a substantial change in electrical conductivity [3] or bleaching in visible absorption [4]. Such chemical modification of electronic structure allows detection of even a single molecule that adsorbs on graphene in a transistor form [5]. The charge exchanges can be described by a simple donor-acceptor model assuming a significant difference in their electron affinities. Despite the apparent simplicity, however, there is a clear lack of understanding in the ubiquitous chemical interactions of 2D materials. The first graphene transistors were reported to be doped with hole carriers originating from unknown chemical entities [6]. Optical confirmation soon followed this observation but with no complete mechanistic understanding [7]. While various acids were exploited to inject hole carriers in graphene, their exchange mechanism is not clear [8]. All of these underline that the spontaneous charge transfer between 2D materials and environment are far from being understood, which hampers both fundamental research and application of 2D materials. In this work, we elucidate that the charge exchange is driven by redox couples of oxygen and water molecules, originally proposed for surface conduction of diamonds [9,10], and that the redox reaction is confined within the nanoscopic space between 2D materials and hydrophilic substrates. By exploiting environment-controlled in-situ spectromicroscopy, we captured molecular diffusion through the interstitial 2D space and revealed that photoluminescence of $WS_2$ and lattice vibration of graphene, both as an indicator of charge density, are dictated by oxygen and water contents and their spatiotemporal distribution.

**Results**

**Chemically modulated ionization of excitons**

Single-layer $WS_2$ ($1LW_{silica}$) was mechanically exfoliated onto $SiO_2/Si$ substrates (see Methods) and served as a model 2D system with precise control over gas environments (Fig. 1a & Supplementary Fig. 1). Photogenerated excitons ($X^0$) have substantial binding energy due to dimensionally restricted screening [11] and interact with excess charge carriers to form charged excitons or trions ($X^\pm$) with additional stabilization as depicted in Fig. 1b [12]. As shown in Fig. 1c, the photoluminescence (PL) spectra of $1LW_{silica}$ in the ambient air typically showed two peaks each originating from excitons ($X^0$ at $E_{X0} = 2.01$ eV) and negative trions ($X^-$ at $E_{X-} = 1.98$ eV), the latter of which derived from defect-related native charge carriers [12]. Assuming that the ionization reactions of excitons into trions are governed by the law of mass action [13], their PL intensity ratio ($I_{X-}/I_{X0}$) serves as a quantitative measure for the density of charge carriers that are modulated by chemical interaction between $WS_2$ and the environments (Supplementary Note 1). When the optical gas cell was purged with an Ar gas, $I_{X-}/I_{X0}$ increased noticeably, which indicated an increase in electron density ($n_e$). The concomitant decrease in the total intensity ($I_t = I_{X-} + I_{X0}$) is due to the increasing importance of the nonradiative decay channel at higher $n_e$ [14]. Water vapor in the Ar gas, however, reduced $n_e$ when judged from a noticeable decrease in $I_{X-}/I_{X0}$. Exposure to dry or humid $O_2$ led to an even more considerable decrease in the ratio and thus further depletion of negative carriers. Time-lapse measurements in Fig. 1d showed that the total PL intensity varied reversibly between Ar and $O_2$ atmospheres, which confirmed the molecular origin of the charge modulation. The dissociation energy of the trions ($E_{diss} = E_{X0} - E_{X-}$) showed a strong correlation with the intensity ratio over the entire atmospheric conditions (Fig. 1e). The overall trend is similar to what was observed in electrically gated measurements [15,16] as shown in Fig. 1e. Since $E_{diss}$ is linearly dependent on the Fermi level [12], smaller $E_{diss}$ for the oxygenic or humid atmosphere indicates significant depletion of electrons induced by hole doping of $WS_2$ by $O_2$ or $H_2O$. Judging from $E_{diss}$ given as a function of $I_{X-}/I_{X0}$ in Fig. 1e, the depletion was near completion and an injection of additional hole carriers would lead to emergence of positively charged trions. A typical difference in $n_e$ was ~$3\times10^{13}$ cm$^{-2}$ between Ar and wet Ar:$O_2$ environments (see Supplementary Note 1 for estimation of charge density). Notably, spatially-resolved spectroscopy showed that the $O_2$-driven rise in $I_t$ was faster at edges than inner areas a few μm off the edges (Supplementary Fig. 2). Moreover, 1LW$_{BN}$ (dry-transferred onto a thin hexagonal BN crystal, hBN) exhibited much smaller $I_t$ and $O_2$-sensitivity than 1LW$_{silica}$ (Fig. 1d). All of these spectral changes indicate that the density of charge carriers is reversibly modulated by a certain charge transfer (CT) reaction that involves $O_2$ and water. The site and substrate-dependent kinetics suggested that the reactions are localized at the interface between $WS_2$ and substrates.

**Interface-confined nanoscopic redox reactions**

Wide-field PL imaging revealed how the CT reaction evolved in space and time as shown in Fig. 2, where 1LW samples were uniformly illuminated with 514 nm laser light under controlled gas or liquid environments. In the ambient air, the intensity image of 1LW$_{silica}$ exhibited slight local irregularities over the entire sample area of ~70 μm$^2$ (Fig. 2a). The multilayer neighbors showed negligible emission due to their indirect bandgaps [17,18]. Because of the O$_2$-driven CT mentioned above, the overall intensity decreased ~50% in an Ar gas and was recovered within 20 minutes under O$_2$ flow, which is consistent with the reversibility shown in Fig. 1d. Notably, however, the recovery in Fig. 2a showed an evident edge-to-center propagation as shown in the enhancement images that are normalized by that of the deoxygenated state in Ar. Exposure for 10 s led to a sharp enhancement by ~30% at the two edges with the other two connected to the multilayers remaining silent. Within 60 s of O$_2$ flow, the two edges exhibited ~100% enhancement, and the enhancement fronts moved further gradually reaching a steady-state after 20 min (Supplementary Movie 1; Supplementary Fig. 3 for more examples) followed by its reversal in subsequent Ar flow (Supplementary Movie 2). In contrast, neither directional enhancement nor complete recovery was observed for 1LW$_{BN}$ (Fig. 2c). On average, the enhancement was small (20 ~ 30%) and occurred rather uniformly, which agrees with Fig. 1d. All of these observations indicated that the CT reactants diffuse through the WS$_2$-silica interface as illustrated by the thick black arrow in Fig. 2d. Noting that the upper surfaces of samples were accessible to the reactants, these results implied a crucial role of the interface or surface of silica and led us to a mechanistic model with two essential ingredients. First, we explain that charge density of WS$_2$ is modulated by the redox reaction involving oxygen and water: $O_2 + 4H^+ + 4e^- \leftrightarrow 2H_2O$ (Equation 1), where reduction of O$_2$ is accompanied by oxidation (hole doping) of WS$_2$ (see Supplementary Note 2 for a detailed account on charge transfer at solid-liquid interface based on Gerischer model). The reaction in Equation 1 was validated for surface transfer doping of hydrogenated diamonds [10], the electrochemical nature of which was first proposed by F. Maier et al.'s [9]. Similar observations were also made for other material systems including GaN [19], ZnO [19], carbon nanotubes [20], and graphene [21]. Second, the CT reaction is localized at the nanoscopic interface by hydrophilic surface functional groups of $SiO_2$ that accommodate water molecules because of the energy gain by hydration of ionic species.

Under this model, the directional propagation of the PL enhancement (Fig. 2a and Supplementary Movies 1 & 2) indicates that the rate-determining step is not the CT reaction but the diffusion of the CT reactants. It is to be noted that the CT rate will decrease with the reaction proceeding according to a simple theory for the CT reaction (Supplementary Note 2). As depicted by the black arrow in Fig. 2d, the diffusion of $O_2$ through $WS_2$-silica interface is initiated from the edges and substantially hindered because of the narrow interfacial gap. The typical propagation rate for $1LW_{silica}$ was ~1 μm/min with a notable variation among samples (Supplementary Fig. 3). On the other hand, the substantially reduced $O_2$-sensitivity of $1LW_{BN}$ (Fig. 1d & Fig. 2c) can be attributed to less hydrophilic nature of hBN that does not hold sufficient water molecules required for the redox reaction (Fig. 2f). In addition, the effective interstitial gap in $1LW_{BN}$ will be much smaller than that in $1LW_{silica}$ because of the flatter substrates and consequently enhanced adhesion for the former, and thus interfacial diffusion of oxygen required for CT will be greatly attenuated. This model is also supported by the substantially enhanced $O_2$-sensitivity and non-directional PL enhancement of $1LW_{silica}$ immersed in water. As shown in Fig. 2b, its PL intensity was increased by ~200% in the presence of dissolved $O_2$, but the enhancement did not show any noticeable spatial propagation (Supplementary Movies 3 & 4). Unlike the gas-phase reaction (Fig. 2d), the top surface of $WS_2$ is in direct contact with water that serves as a hydrating solvent and thus works as a major CT route as illustrated by the thick blue arrow in Fig. 2e.

**Boosting and quenching of redox reactions**

To unveil the pivotal role of hydrophilic $SiO_2$ surface in the redox reaction, we exploited charge density-dependent phonon hardening of graphene. As shown in Fig. 3a, G and 2D peaks ($\omega_G$, $\omega_{2D}$) of 1L graphene/$SiO_2$/Si ($1LG_{silica}$) were greatly upshifted and recovered by thermal annealing at 500 and 1000 °C, respectively. Based on the established Raman metrology [22] of hole density ($n_h$) and lattice strain ($\varepsilon$) shown in Fig. 3b, $\omega_G$ and $\omega_{2D}$ of samples that were treated at various temperatures were translated into $n_h$ and $\varepsilon$ with high accuracy of ~$1\times10^{12}$ cm$^{-2}$ and ~ 0.2%, respectively (Fig. 3c). $\Delta n_h$ reached a maximum of 1.0 ~ $1.5\times10^{13}$ cm$^{-2}$ with slight changes in strain when vacuum-annealed at 400 ~ 700 °C (Fig. 3c). Despite many studies [23-27], however, the mechanistic origin of the thermally activated CT or hole doping has been unclear. In the following, we propose and show that it is the same redox reaction as Equation 1 and thermal activation renders $SiO_2$ surface hydrophilic enough to bind a certain amount of water molecules. As illustrated in Fig. 3e (left), the surface of silica is terminated with hydrophobic siloxane (Si-O-Si) and hydrophilic silanols (Si-OH). At elevated temperature, the latter transforms into the former without the presence of water (dehydroxylation) and vice versa with water (hydroxylation) [28]. When annealed at < ~750 °C, the $SiO_2$ surface becomes hydroxylated at the expense of water trapped during sample preparation. When placed in the ambient air, more water molecules are attracted at the graphene-$SiO_2$ interface as depicted in Fig. 3e (middle) and electrons of graphene are consumed by the interfacial redox reaction leaving graphene highly hole-doped. When treated at > ~750 °C, however, dehydroxylation was more dominant as shown in Fig. 3e (right) and $\Delta n_h$ dropped to 1/3 of the maximum (Fig. 3b & 3c). Competition between hydroxylation and dehydroxylation was verified by measuring water contact angles (WCA) for bare substrates (see Supplementary Fig. 4 for raw data). As shown in Fig. 3c, $\Delta n_h$ has a clear correlation with WCA and thus hydrophilicity of substrate surfaces. To further corroborate the hypothesis, we prepared samples in a glove box with interfacial water further minimized by pretreating substrates with diethyl zinc vapor that removes even trace-amount of surface water [29]. The charge-density maps obtained after annealing at 400 °C showed that CT is negligible for diethyl zinc-treated samples unlike non-treated samples (Fig. 3d). This control experiment confirmed that interfacial water is required for the thermal hydroxylation which leads to the activated CT.

**pH-controlled redox reactions**

We now show that the interfacial CT reaction can be generalized to a pH-dependent redox between 2D materials and $O_2$ dissolved in liquid water. High-purity $O_2$ or Ar gas was sparged through an optical liquid cell containing HCl solution of a preset pH to control the concentration of dissolved $O_2$, $[O_2]$ (Fig. 4d). When $O_2$ was introduced to the Ar-saturated HCl solution of pH = 2, the G and 2D peaks of $1LG_{silica}$ upshifted (Supplementary Fig. 5) indicating a significant level of hole doping ($\Delta n_h \sim 4.0 \times 10^{12}$ cm$^{-2}$) (Fig. 4a). Notably, $n_h$ decreased reversibly with Ar gas bubbled through the solution, and the doping-undoping cycle could be repeated multiple times. The $O_2$-mediated CT was observed at pH = 1 ~ 4 but not at pH ≥ 5 (Supplementary Fig. 5). The rise and decay kinetics of $n_h$ are highly pH-dependent and self-limited (Fig. 4b & 4c). A typical initial CT rate is $\sim 1 \times 10^{10}$ cm$^{-2}$s$^{-1}$ at pH = 2 and $[O_2]$ = 1.3 mM (see Methods). As shown in Fig. 2b, PL signals of $WS_2$ also exhibited the same sensitivity towards dissolved $O_2$ in water, and the change became more obvious at lower pH (Fig. 4f & Supplementary Fig. 6). Equation 1 shows that one oxygen molecule may exchange 4 electrons with other materials in the presence of protons. According to the Gerischer model on CT at an electrode-liquid interface [30], the direction of CT is determined by energetic alignment between the Fermi level ($E_F$) of 2D materials and electrochemical potential ($E_{F,redox}$) of the redox system as shown in Fig. 4e (Supplementary Note 2). Since $E_{F,graphene}$ = -4.57 eV and $E_{F,redox}$ = -5.669 + 0.0592pH - 0.0148log[p($O_2$)], reduction of $O_2$ is more favored at lower pH, which leads to increased hole doping of graphene (Supplementary Text B). On the other hand, the CT is doubly inhibited by a reduction of the reactant and an increase in $E_{F,redox}$, when the concentration of $O_2$ is decreased. The rate of CT is proportional to the density overlap between occupied states of graphene and empty oxidized states of the redox system, the latter of which is depicted as a Gaussian distribution centered at $E_{ox}$ in Fig 4E. The self-limited CT observed even in the presence of sufficient reactants (Fig. 4b) is due to the CT-induced decrease in $E_{F,graphene}$, and can be well described by the Gerischer model (Supplementary Note 2). For $WS_2$ that is natively n-doped by various defects including S vacancies [31], its Fermi level is near the conduction band minimum located at -3.93 eV [32] that is 0.64 eV higher than that of graphene. Thus, CT from $WS_2$ to the redox couples of $O_2/H_2O$ is more favorable than from graphene, which is consistent with our finding that the pH threshold for $O_2$-induced CT is higher for $WS_2$ than that for graphene.

**Discussion**

In this work, we have shown that the ambient oxygen reduction reaction is behind the longstanding mystery of the spontaneous and activated charge transfer doping in graphene and $WS_2$. Redox couples of $O_2/H_2O$ responsible for the charge transfer reside at the interface of 2D materials and substrates, and their 2D diffusion was captured in real-time by wide-field photoluminescence imaging. The CT reaction can be turned on and off via controlling $O_2$ or interfacial water serving as a hydration solvent. The CT rate of 2D materials in contact with liquid water can also be tuned by varying the concentration of dissolved protons or $O_2$ as described by the Nernst equation. The presented nanoscopic electrochemistry will pave the way towards efficient control of charge density and related material properties in 2D and other low-dimensional materials and devices.

## Methods

### Preparation and treatment of samples

Most of single-layer WS$_2$ (1LW$_{silica}$) and graphene (1LG$_{silica}$) samples were prepared by mechanically exfoliating bulk crystals (WS$_2$ from 2D semiconductors Inc.; graphite from Covalent Materials Inc. and Naturgaphit GmbH) onto SiO$_2$ (285 nm)/Si substrates in the ambient environment. For 1LW$_{BN}$ samples, single layers were first mechanically exfoliated onto polydimethylsiloxane (PDMS) substrates and then dry-transferred onto thin hBN crystals supported on SiO$_2$/Si substrates. To minimize a change in spectroscopic signals induced by optical interference, hBN crystals thinner than 3 nm were selected as substrates. To remove interfacial water in 1LG$_{silica}$, we exposed bare substrates briefly to the vapor of diethyl zinc before mechanical exfoliation in a glove box. For thermal activation, samples were annealed at a target temperature for 2 hours in a quartz tube furnace that was maintained at a pressure of 3 mTorr.

### Raman and photoluminescence (PL) measurements

Raman and photoluminescence (PL) were performed with a home-built micro-Raman spectrometer setup [22]. Briefly, monochromatic outputs from solid-state lasers operated at 458 nm and 514 nm were focused onto samples with a spot size of ∼1 μm using a microscope objective (40X, numerical aperture = 0.60). Backscattered PL and Raman signals were collected with the same objective and guided to a spectrometer equipped with a liquid nitrogen-cooled CCD detector. Overall spectral accuracy was better than 5 and 1 $cm^{-1}$ for PL and Raman measurements, respectively. To avoid any significant photo-induced effects, we maintained the average power of the excitation beam below 6 μW for $WS_2$ and 400 μW for graphene samples.

For wide-field PL imaging, the collimated green laser beam was focused at the back-focal plane of the objective with a plano-convex lens (focal length = 400 mm) after 3-times expansion with a Galilean beam expander. The average power of the wide-field excitation was maintained below 1.5 mW that was illuminated onto an area with a diameter of ~100 μm. PL signals in the range between 1.9 and 2.1 eV mostly contributed to the PL images recorded with the CCD detector.

**Control of gas and liquid environments**

For optical measurements in controlled gas environments, samples were mounted in a custom-made optical gas cell with precise controls over the flow rates of Ar (20 ~ 1000 mL/min) and $O_2$ (5 ~ 250 mL/min). Unless otherwise noted, flow rates for dry gases were 1000 and 250 mL/min for Ar and $O_2$, respectively. Relative humidity (RH) inside the gas cell was monitored by a serially connected hygrometer and could be controlled between 5 and 90% by varying the mixing ratio of an additional Ar gas line that passed through a water bubbler (Supplementary Fig. 1a). The temporal change of RH upon injection of wet gas could be well fitted with two exponential functions with time constants of 15 ± 2 and 116 ± 40 s. The fast component is limited by the finite flow rate through the gas manifold, and the slower one is related to adsorption kinetics of water on the inner walls of the manifold and gas cell. The former time constant set an upper bound for the average arrival time for the dry gas experiments since the gas manifold was smaller for the dry gas case. The equilibrium RH values were in a linear relation with the mixing ratio of the wet Ar gas (Supplementary Fig. 1b), showing good controllability.

For in-situ measurements in aqueous solutions, samples were placed in a custom-designed Teflon-based optical liquid cell with a gas sparger. The concentration of dissolved oxygen was controlled by flowing high-purity $O_2$ or Ar gases at 250 mL/min through the sparger, which is an efficient method to saturate or eliminate dissolved oxygen [33]. When saturated under 1 atm $O_2$ gas, a simple estimation using Henry's law at 298 K predicts that the concentration of $O_2$ will reach 1.3 mM. On the other hand, a 5-min or longer sparging of inert gas leads to a residual concentration of ~15 μM or less [33]. pH of aqueous solutions was varied with hydrochloric acid.

**Contact angle measurements**

The hydrophilicity of $SiO_2$/Si substrates was quantified by water contact angles to reveal competition between thermally activated surface hydroxylation and dehydroxylation [28] in the presence of water vapor. Bare $SiO_2$/Si substrates were annealed at various temperatures for 2 hours in the tube furnace filled with water vapor and $O_2$ gases. Within 20 min after the treatments, water contact angles of the substrates were measured in the static mode with an optical tensiometer (SmartDrop, Femtobiomed Inc.) in ambient conditions. The volume of water droplets was maintained in the range of 2.5 ~ 3 μL.

## Data availability

The data that support the findings of this study are available in the Supplementary Information and from the corresponding author upon reasonable request.

**Acknowledgments:** We thank Dong-Gyu Lee and In Su Lee for assistance in preparing samples using a glove box. This work was supported by Samsung Research Funding Center of Samsung Electronics under Project Number SSTF-BA1702-08.

**Author contributions:** S.R. proposed and directed the research; K.P., H.K., S.K. and D.L. carried out experiments and data analysis; S.R., K.P. and H.K. wrote the manuscript; and all authors contributed to discussions.

**Competing interests:** We declare no competing interests.

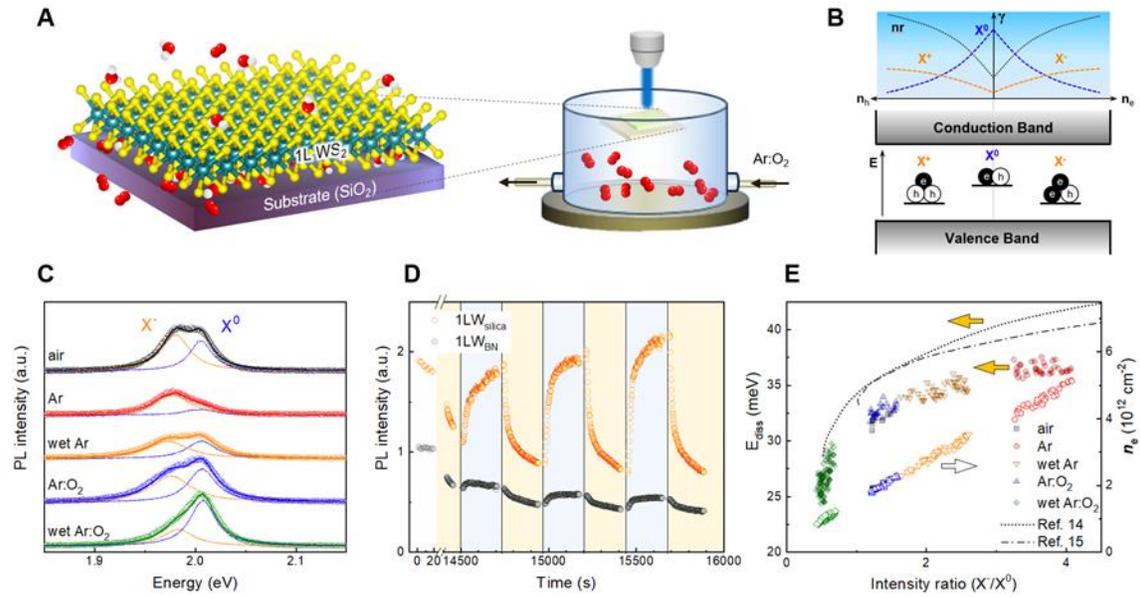

**Fig. 1** Chemically modulated ionization of excitons. **a** In-situ optical measurements of single-layer (1L) $WS_2$ supported on $SiO_2$ ($1LW_{silica}$) located in a controlled gaseous environment. The scheme illustrates $1LW_{silica}$ interacting with $O_2$ and water in a redox reaction. **b** Electronic energy bands of $WS_2$ with neutral ($X^0$) and charged ($X^+$ and $X^-$) excitonic states (bottom panel); Schematic representation of their radiative and non-radiative (nr) decay rates ($\gamma$) given as a function of charge density ($n_e$ and $n_h$) (Ref. 14) (upper panel). **c** Photoluminescence (PL) spectra of $WS_2$ in various gas environments. Solid lines are double Lorentzian fits representing $X^0$ and $X^-$. Relative humidity for wet gases was 45%. **d** Time-lapse measurements of PL intensity of $1LW_{silica}$ and $1LW_{BN}$ in response to changes in gas environments: ambient air (white box), Ar (yellow box) and $Ar:O_2$ = 4:1 (blue box). **e** Equivalence between chemical (filled symbols) and electrical (lines, Refs. 15 & 16) modulations of charge density in $WS_2$, where trion dissociation energy ($E_{diss}$) in the left ordinate was given as a function of PL intensity ratio ($X^-/X^0$). The electron density ($n_e$) in the right ordinate (open symbols) was determined for the filled symbols according to the mass action law for excitons and trions (Supplementary Note 1).

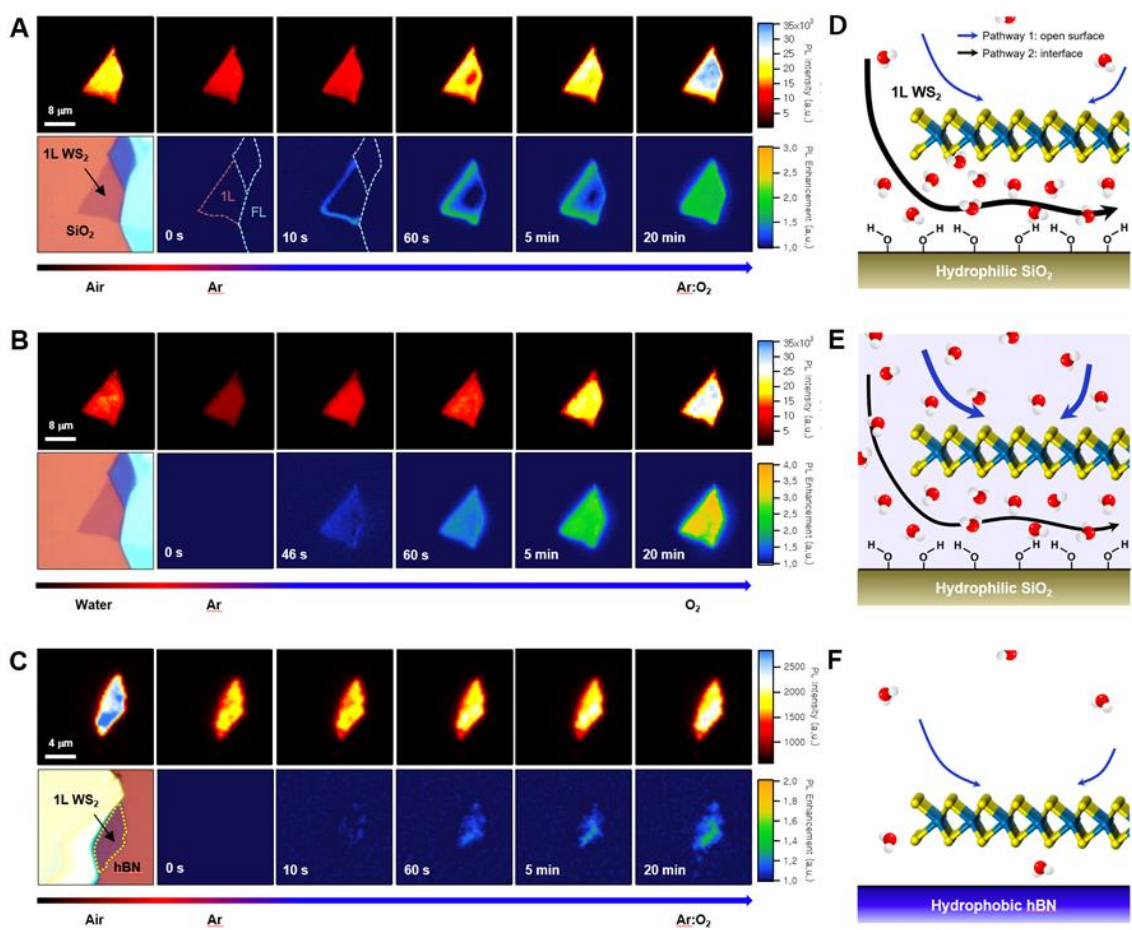

**Fig. 2** Real-time photoluminescence images of interface-confined redox reactions. **a~c** Wide-field PL images (top rows) and optical micrographs followed by PL enhancement images (bottom rows) of 1LW$_{silica}$ in gas (**a**) and water (**b**), and 1LW$_{BN}$ in gas (**c**) environments. Outlines of single and few-layer (FL) areas (dashed lines) were given in the first two enhancement images of **a**. Exposure time was 1.5 s for each image. The enhancement images were obtained by dividing PL images with that for time zero. Samples in **a~c** were pre-equilibrated with Ar gas for 2 hr before exposure to Ar:O$_2$ mixed gas (O$_2$ gas for **b**) at time zero. For **b**, gases were bubbled through a sparger immersed in the optical liquid cell. Scale bars: 8 (**a** & **b**) and 4 (**c**) μm. **d~f** Schemes for major diffusion routes of O$_2$ responsible for redox reactions of 1LW$_{silica}$ in the gas phase (**d**), 1LW$_{silica}$ in water (**e**), and 1LW$_{BN}$ in the gas phase (**f**). The thickness of each arrow represents the relative contribution of each pathway to overall charge transfer reactions.

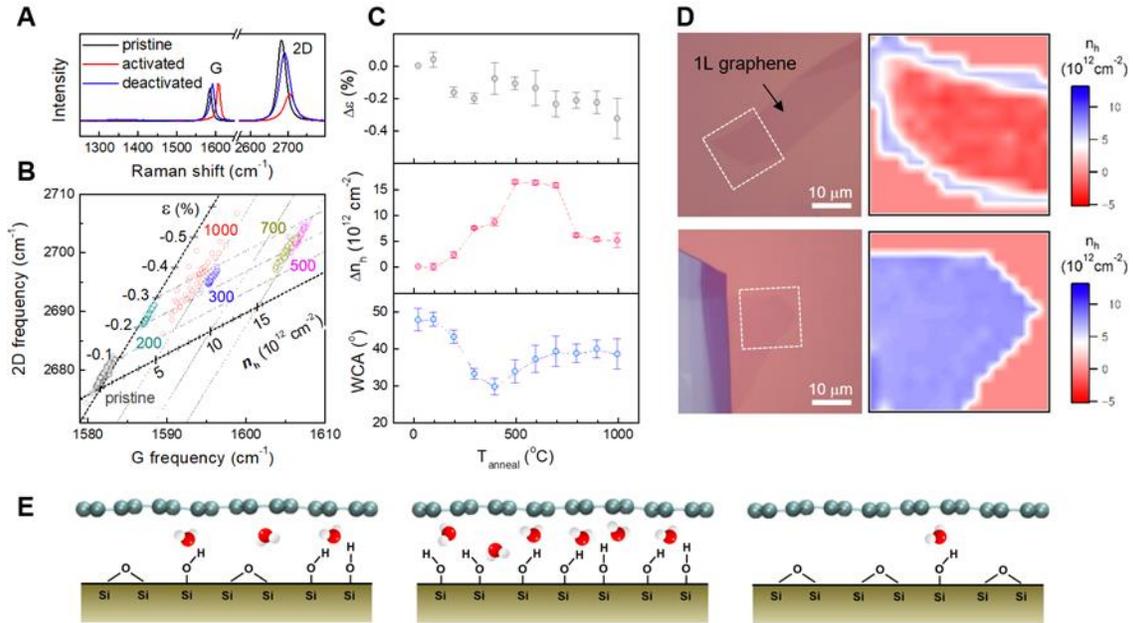

**Fig. 3** Amplification and suppression of redox reactions. **a** Raman spectra of pristine (black), thermally activated (red; annealed at 500 °C) and deactivated (blue; annealed at 1000 °C) 1L graphene (1LG$_{silica}$). **b** Correlation between G and 2D frequencies of 1LG$_{silica}$ samples annealed at various temperatures (in °C; given next to data). Lattice strain ($\varepsilon$) and electrical hole density ($n_h$) were determined according to Ref. 22. **c** Changes in lattice strain (top) and hole density (middle) graphene; water contact angle (bottom) of SiO$_2$ substrates as a function of annealing temperature (T). Error bars denote standard deviation. **d** Optical micrographs (left) and charge density images (right) of 1LG$_{silica}$ samples prepared on diethyl zinc-treated (top) and non-treated (bottom) substrates. Both samples were activated by annealing at 400 °C. Scale bars: 10 μm. **e** Schematic representation of air-equilibrated graphene-SiO$_2$ interface: pristine (left), hydroxylated by thermal activation (middle), and dehydroxylated by thermal deactivation (right). Equilibrium coverage of interfacial water is determined by effective hydrophilicity of SiO$_2$ substrates.

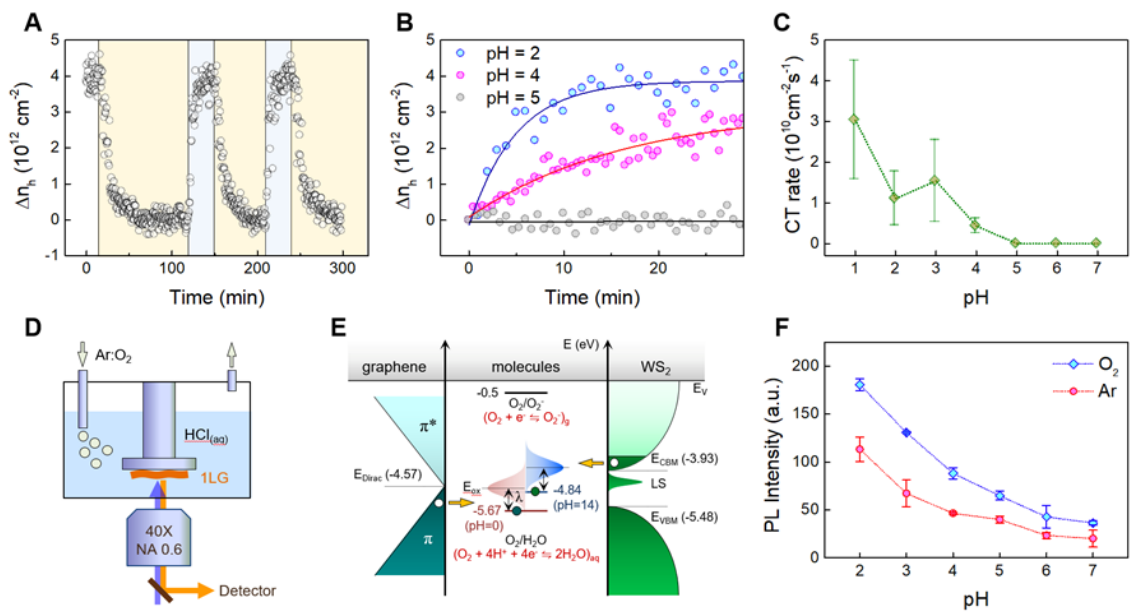

**Fig. 4** pH-controlled charge transfer. **a** Time-lapse measurements of charge density in $1LG_{silica}$ in HCl solution (pH = 2) through which Ar (yellow box) and $O_2$ (blue box) gases were sparged alternatively. The data near time zero (white box) were obtained with the HCl solution aerated. $\Delta n_h$ was referenced to $n_h$ for the Ar flow. **b** pH-dependent kinetics of $O_2$-induced rise in charge density of $1LG_{silica}$ in HCl solutions: pH = 2 (blue), 4 (magenta), and 5 (gray). The solid lines are exponential fit to the data. **c** Initial charge transfer rate per unit area of $1LG_{silica}$ in HCl solutions of varying pH. **d** Scheme of the optical liquid cell with a gas sparger combined to a microspectroscopy setup. **e** Energy level diagram for redox-governed charge transfer from graphene and $WS_2$ to $O_2/H_2O$ redox couples. Electron-accepting oxidized states are represented by Gaussian distributions displaced from $E_{F,redox}$ by a solvent reorganization energy ($\lambda$). The Fermi levels of electron donors are at $E_{Dirac}$ for graphene and near the conduction band minimum (CBM) for $WS_2$. LS represents localized mid-gap states originating from defects. **f** Total PL intensity of $1LW_{silica}$ modulated by $O_2$ dissolved in HCl solution of various pH. Error bars denote standard deviation.